\documentclass[%
 reprint,
 prl,
 amsmath,amssymb,
 aps,
]{revtex4-2}

\usepackage{extarrows}
\usepackage{graphicx}
\usepackage{dcolumn}
\usepackage{bm}
\usepackage{xcolor}

\mathchardef\mhyphen="2D

\begin{document}

	\title{Non random behavior in the projection of random bipartite networks}

    \author{Izat B. Baybusinov}
    \author{Enrico Maria Fenoaltea}%
    \author{Yi-Cheng Zhang}
    \email{yi-cheng.zhang@unifr.ch}
    \affiliation{Physics Department, University of Fribourg, Chemin du Mus\'ee 3, 1700 Fribourg, Switzerland}%
    
    \date{\today}

    \begin{abstract}
There are two main categories of networks that are investigated in the complexity physics community: monopartite and bipartite networks. In this letter, we report a general finding between these two classes. If a random bipartite network is projected into a monopartite network, under quite general conditions, we obtain a non-random monopartite network with special features. We believe this finding is very general and has important real-world implications.   
    \end{abstract}
    
    \maketitle

In the last three decades, networks have been a major research subject\cite{albert2002statistical,boccaletti2006complex,mata2020complex}. Among them, bipartite networks are one particular class \cite{newman2002random,barber2007modularity, ramasco2004self, holme2003network}.
We can present a bipartite network as people connected to events. For example, an agent can participate in a few among many events and if everyone randomly chooses a few events independently, we obtain a typical random bipartite graph. Traditional networks, instead, belong to the so-called class of monopartite networks. We can present a random monopartite network as formed by people only. If each agent chooses randomly to connect with a few other agents, then we obtain a random monopartite network with nodes solely represented by people. This is the well-known Erdos-Renyi network \cite{10.1214/aoms/1177706098}.

Networks of the latter class can be constructed from those of the former class by projection \cite{zhou2007bipartite, cimini2022meta}: A new monopartite network can be constructed where two individuals are connected if, in the bipartite counterpart, they have at least one common event. 
In the literature, the people events network is called affiliation network \cite{faust1997centrality,gallos2012people}, and the projection is studied in many different systems. For instance, scientists connected by collaborating on the same project \cite{sun2011co,uddin2013network}, movie actors connected by appearing in the same movie \cite{amaral2000classes, watts1998collective}, and so forth. In addition, the projected network is the basis of recommender systems \cite{lu2012recommender,resnick1997recommender}.


In this letter, we study the properties of the projected network in the general case of a random bipartite network. 
Similar studies \cite{vasques2020transitivity,guillaume2006bipartite} have examined the projected network by randomizing a configuration model \cite{squartini2011analytical} with a given degree sequence. Here we show that in the most simple approximation, i.e., a random bipartite graph, the projected network has interesting features.

Keeping with the people and events metaphor for bipartite networks, we construct a random network consisting of $K$ people and $N$ events and assume that each person connects to any of the events with probability $\beta \in [0,1]$. In a large sample size, the decision to participate in an event can be assumed to be random.
The elements of the adjacency matrix $\mathbf{n}$ of the bipartite network is written as
\begin{equation}
    n_{i\alpha}= \begin{cases}
    1 \text{ with probability }\beta \\
    0 \text{ with probability }1-\beta, 
\end{cases}
\end{equation}
where individuals are labeled by $i=1,..,K$ and events by $\alpha=1,..,N$.
From this adjacency matrix, we can compute all the network observables. For our purposes, it is useful to work with the degree distribution, i.e., the probability $B_N(m)$ that an individual is connected to $m$ events out of $N$. Since the network is random, it follows a binomial distribution: 
\begin{equation}
B_N(m)= \binom{N}{m}\beta^m(1-\beta)^{N-m}.
\end{equation}
Now, we focus on the network of people resulting from the projection of the bipartite network. Since two agents are connected when they share at least one event, the elements of the adjacency matrix $\mathbf{A}$ in the projected monopartite network can be written as:
\begin{equation}
A_{ij} = \theta\left(\sum_{\alpha=1}^N n_{i\alpha}n_{j\alpha} \right), 
\label{A}
\end{equation}
where $\theta(\cdot)$ is the Heaviside function, and $i,j=1,..,K$.

We want to study the properties of $\mathbf{A}$ averaged over all the realizations of the bipartite network. However, writing the probability $\mathbf{A}$ involves $K^2$ conditions on the adjacency matrix. Therefore, we compute a simpler quantity, that is, the probability $p$ that an element $A_{ij}$ is 1. For $N$ events this is the extremal distribution of sampling at least 1 element out of $N$, and correlations due to the transition from a bipartite to a monopartite structure are averaged out. We have

\begin{equation}
p= \Biggl\langle\theta\left(\sum_{\alpha} n_{i\alpha}n_{j\alpha} \right)\Biggl\rangle =  1-(1-\beta^2)^N
\end{equation}
By symmetry, this is the same for any $i$ and $j$, so the link probability gives no information about the bipartite structure in the monopartite version, and one might think that the network is random if we neglect correlations.
In the following, we compute $P(d)$ of the degree distribution of the projected network. The degree $d_i$ of the individual $i$ in the monopartite network is given by $d_i=\sum_{j}A_{ij}$. 
If an individual with degree $d$ participates in $m$ events, there must be $d$ individuals out of $K-1$ (we do not allow self-link) meeting her at one of those $m$ events. Summing over $m$ we have
\begin{equation} \label{P(d)}
 P(d) = \sum_{m=0}^N B_N(m)  \binom{K-1}{d} [1-B_m(0)]^{d}B_m(0)^{K-1-d}.
\end{equation}
When $N\to\infty$ and $\beta$ is fixed, the sum in Eq.\ref{P(d)} is dominated by the maximum of $B_N(m)$, i.e., when $m=\beta N$. In this limit, $P(d)$ is a binomial distribution with the probability parameter given by $1-e^{-\beta^2N} \approx p$. Thus, one retrieves the Erdos-Reny network with connection probability $p$. 

In general, one can calculate the moment-generating function to show that $P(d)$ is binomial in the large $N$ limit:
\begin{equation}
\begin{split}
    \left\langle e^{\lambda d} \right\rangle = e^{\lambda (K-1)}
    \sum_{l=0}^{K-1} 
    &\biggl[\left(\beta(1-\beta)^l+(1-\beta)\right)^N\\
    &\times  \binom{K-1}{l} (e^{-\lambda}-1)^l  {\biggr]} ,
   \end{split}
   \label{mgf-bipartite}
\end{equation}
and at large $N$ we have $\left[\beta(1-\beta)^l+(1-\beta)\right]^N\approx e^{-l\beta^2N}$. So we obtain
\begin{equation}
    \left\langle e^{\lambda d_i} \right\rangle \approx e^{\lambda (K-1)} \left[ 1 - e^{-\beta^2N}(1-e^{-\lambda}) \right]^{K-1},
\end{equation}
which correspond to the moment-generating function of a binomial distribution with probability parameter $1-e^{-\beta^2N}$. Hence, with many events, the projected network and a random monopartite network cannot be distinguished when measuring their degree distributions. Note that the above argument fails if $K$ scales with $N$.

\begin{figure}
    \centering
    \includegraphics{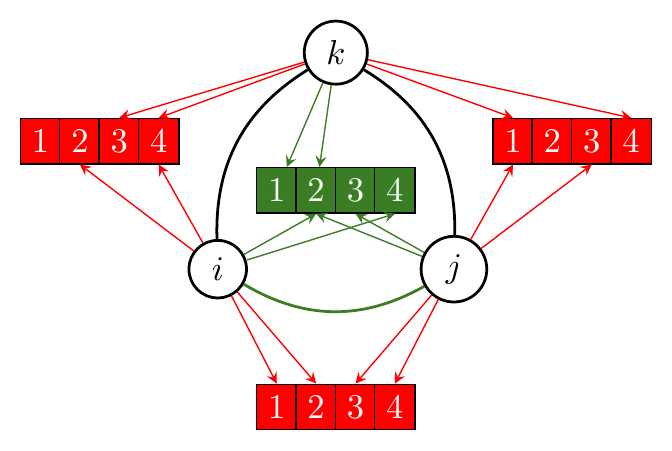}
    \caption{Difference between a projected bipartite network and a random network (i.e., the projected network without considering correlations) with 4 events and 3 agents. In the bipartite configuration, there is a unique list of events (in green). Meanwhile, in the random network case, there is an independent list of events (in red) for each possible link, i.e., for each pair of individuals.}
    \label{fig:1}
\end{figure}

\begin{figure}[ht]
    \centering
    \includegraphics[scale=.8]{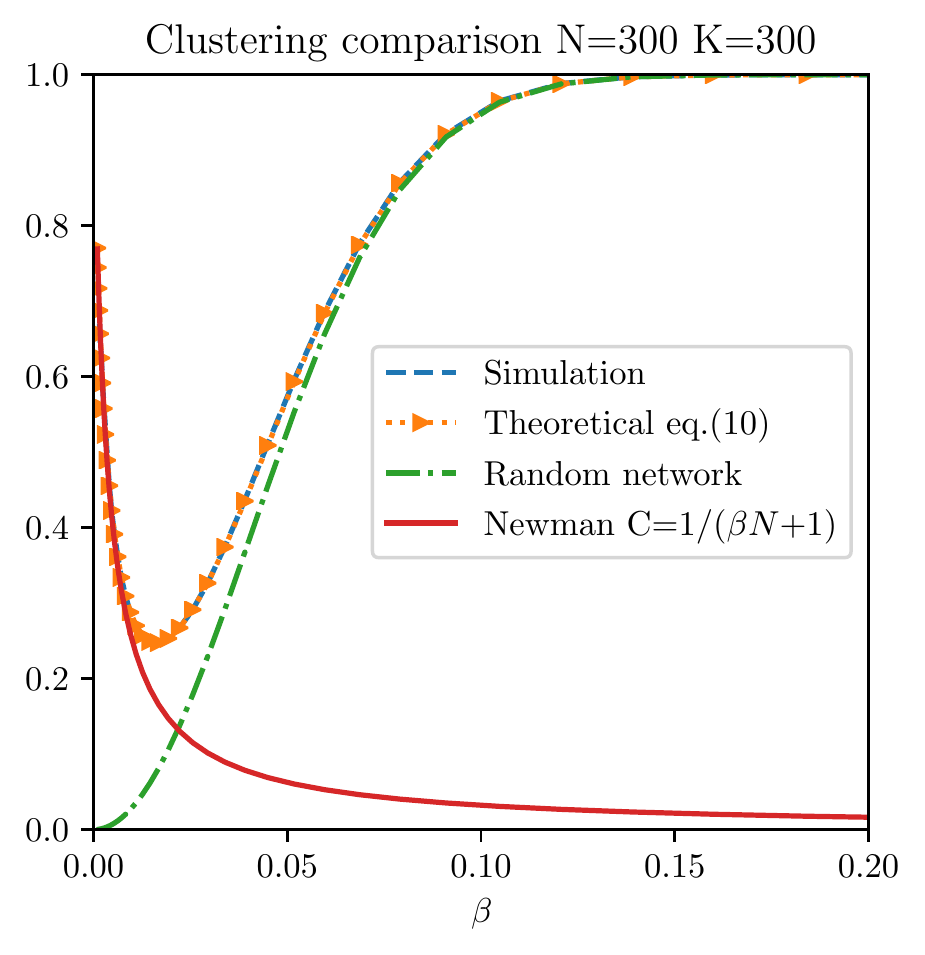}
    \caption{Clustering coefficient vs $\beta$. The clustering of our projected network is compared with that of a random network with connection probability $p$ and that of the Newman model in \cite{newman2002random}.}
    \label{fig:plots}
\end{figure}
\begin{figure}[ht]
    \centering
    \includegraphics[scale=.8]{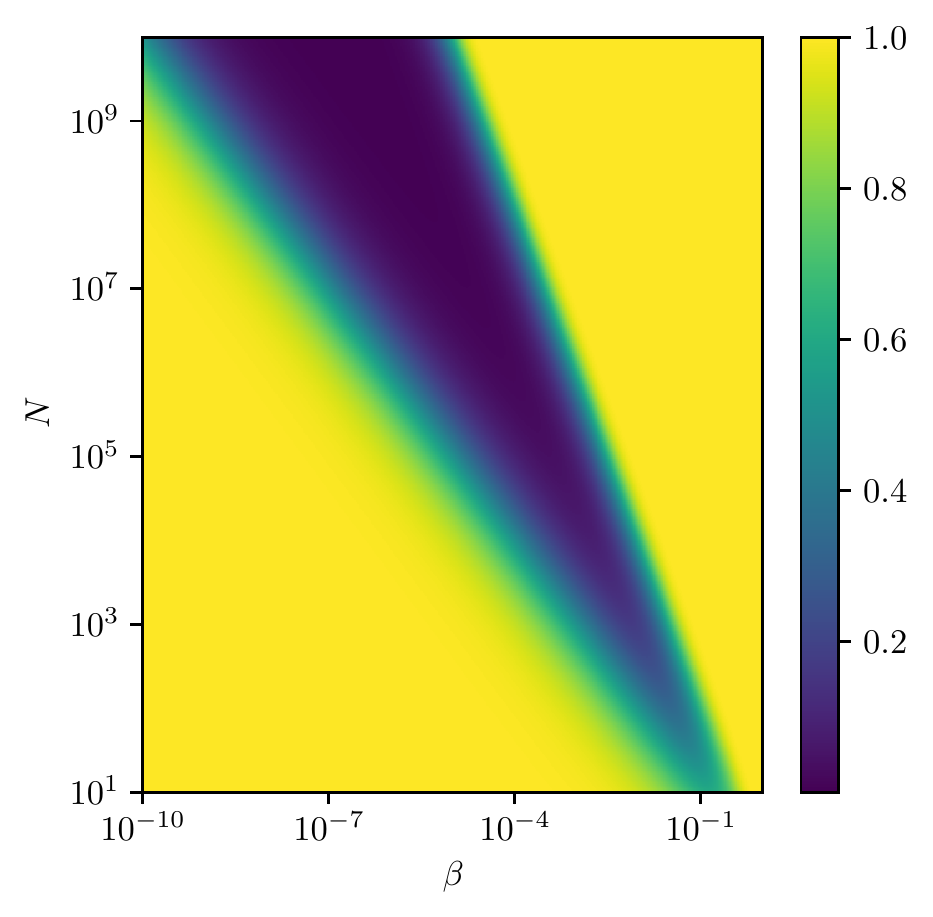}
    \caption{Phase diagram of the clustering coefficient of Eq.\ref{eq:C} in the $[\beta,N]$ parameter space. When $N$ increases, the width of the minimum valley increases since the left border scales as $1/N$ and the right one as $1/\sqrt{N}$.}
    \label{fig:diagram}
\end{figure}

Besides the degree distribution, we also consider a higher-order measure, namely,  the clustering coefficient $C$ of the projected network \cite{watts1998collective}:
\begin{equation}
    C = 3\frac{\langle{ triplets }\rangle}{\langle{open\,\, triplets}\rangle } =
    \frac{\sum_{i,j,k}\langle{A_{ij}A_{jk}A_{ki}}\rangle}{ \sum_{i} \langle{(\sum_{j}A_{ij})(\sum_{j}A_{ij}-1) }\rangle }
    \label{C}
\end{equation}
This quantity measures how connected a node's neighbors are to one another. Note that, in the case of a random network generated with probability $p$, the clustering coefficient is $p$ \cite{li2017clustering}. To compute the numerator of Eq.\ref{C} in our projected network, we must consider the probability that three agents $i,j,k$ are connected. There are two contributions to this probability. The first is the probability $P_2$ that in the bipartite network $i,j,k$ participate in the same event. In this case, we have $P_2=1-(1-\beta^3)^N$. The second is the probability $P_3$ that the three individuals participate pairwise in three different events. This is computed as follows:
\begin{equation}
    P_3 = \sum_{m,n,l} B_N(m) B_m(n) \left[ B_n(0) B_{m-n}(l)\right] (1-B_{N-m}(0))
\end{equation}
The first factor in this sum is the probability that the first individual, say $i$, participates in $m$ events out of $N$; the second and third factors are the conditions that $j$ participates in $n$ of these $m$ events and that $k$ participates in $l$ of the remaining $m-n$ events, respectively. In this way, individuals $j$ and $k$ both meet $i$, but they do not meet each other in any of the $m$ events chosen by $i$. The last factor is the condition that $j$ and $k$ meet in the other $N-m$ events. The denominator of Eq.\ref{C} is straightforwardly written as $ K \langle{d_i(d_i-1)}\rangle$. After some algebra, the clustering coefficient can be written as:
\begin{equation}
C =3-\frac{2+(1+2\beta)^N(1-\beta)^{2N}-3(1-\beta^2)^N}{1-2(1-\beta^2)^N+(1-2\beta^2+\beta^3)^N}.
\label{eq:C}
\end{equation}
Fig.\ref{fig:plots} shows that our analytical result is consistent with numerical simulations.
The clustering is 1 at $\beta\approx 1/N$, where the network is separated into disconnected communities composed of individuals that share all the events they participate in (i.e., in a community two people participate in the same events and do not participate in all the others). Increasing $\beta$, spurious links start to appear between communities, reducing the clustering until $\beta \sim N^{-1/2}$ (see Fig.\ref{fig:diagram}), where links between two randomly sampled individuals are highly frequent.
From this point on, the increase in link density generates a higher clustering coefficient. Actually, the clustering reaches its minimum value when the network transits from a phase of isolated communities to one with a unique connected component. To evaluate this minimum we compute $C$ for small $\beta$:
\begin{equation}
    C=  \frac{1+N^2\beta^3 }{N \beta +1}+o(\beta^3).
\end{equation}
The minimum is obtained by solving the following equation:
\begin{equation}
2\beta^3 N^3 +3\beta^2N^2 - N = 0.
\end{equation}
For large $N$, the solution $\beta_c$ scales as $N^{-2/3}$. Instead, the clustering evaluated at $\beta_c$ decreases quite slowly with $N$:
\begin{equation}
C(\beta_c) \approx N^{-1/3}.
\end{equation}
Note that, when looking at the clustering coefficient, our projected network is fundamentally different from a random network, as shown in Fig.\ref{fig:plots}. Unlike the case of the degree distribution, this is also true for large $N$. This is a consequence of the correlations arising from dimensionality reduction. Therefore in the projection of a random network, the degree distribution is not a sufficient measure to extract all the information from a network.

The presence of correlations can be explained by a geometrical argument. For any individual $i$, let us define $\mathbf{v}_i=\{n_{i\alpha}\}_{\alpha=1,..,N}$ as the vector whose elements are 1 if $i$ participates in event $\alpha$, and zero otherwise. We can interpret this vector also as a vertex of an $N$-dimensional hypercube. Thus, each vector (or individual) is a randomly drawn vertex on the hypercube with an average distance from the origin equal to $\beta N$. On this hypercube two individuals are connected whenever their inner product is positive, that is, when $\mathbf{v}_i \cdot \mathbf{v}_j>0$. From this perspective, correlations in the projected network naturally appear as dimensional constraints on the hypercube.

Now, it is interesting to study the properties of our system in function of $N$ as the scaling gives different behaviors. For example, in \cite{newman2002random}, Newman describes a similar model where he fixes the average degree $z=O(1)$ in the bipartite network so that $\beta = z/N$. Since $\beta$ scale as $1/N$, for large $N$, the resulting monopartite network of people is clustered into different communities, as we have shown above. 
Thus, with the Newman approach, the effects arising at higher values of $\beta$ cannot be observed (see Fig.\ref{fig:plots}), and the network always remains fragmented.

Remarkably, the number of events $N$ can be used as a scaling tool of the network: studying the system with few events (for example, by aggregating similar events or by limiting the data set) lead to a richer scenario where links between different communities appear. One can argue that interesting social phenomena can be detected by observing the system at a fixed scale (in this case $N$).

In the real world there is a cognitive constraint in the number of connections each individual can handle \cite{dunbar1992neocortex}. 
To take this into account, we can fix the average degree in the projected network to be of the order of 1:
\begin{equation}
\langle d_i \rangle = K p = K[1- (1-\beta^2)^N]\sim O(1),
\end{equation}
for any $i$. In this way, we obtain a new scaling:
\begin{equation} 
1- (1-\beta^2)^N \sim 1/K \quad \rightarrow \quad     \beta\sim \sqrt{\frac{1}{NK}}
\end{equation}
So the probability to participate in an event depends on both $N$ and $K$. If $K \approx \sqrt{N}$ the population is clustered in different communities. Instead, if $K < \sqrt{N}$, the system is more cohesive. This mechanism can help to understand and manage social phenomena such as the observed social fragmentation and polarization \cite{minh2020effect, pham2021balance}. 

To generalize these results from a statistical mechanics perspective, one can introduce a Hamiltonian for the matrix $\mathbf{A}'$:
\begin{equation}
    H[\mathbf{A}'] = \sum_{i=1}^K\sum_{j=1}^K A_{ij}' (1-\mathbf{v}_i \cdot \mathbf{v}_j).
\end{equation}
This describes a model where individuals have a nonzero probability (defined by the temperature) of being connected in the projected network even though they have no common events in the bipartite network. The minimum of the Hamiltonian is when $\mathbf{A}'$ is identically the adjacency matrix in Eq.\ref{A}. So, by studying this Hamiltonian in the $0$ temperature limit, one can study the properties of $\mathbf{A}$. Computing the partition function $Z$, we find
\begin{equation}
    Z = \prod_{ij}\sum_{A_{ij}'=0}^1 e^{-\beta H[\mathbf{A'}]} =  \prod_{ij}\left(1+e^{-\beta (1-\mathbf{v}_i \cdot \mathbf{v}_j)}
    \right),
\end{equation}
and the link probability is:
\begin{equation}
    n_{ij}:=\sum_{A_{ij}=0}^1A_{ij}P(\mathbf A_{ij}) = \frac{1}{e^{\beta (1-\mathbf{v}_i \cdot \mathbf{v}_j)} +1 },
\end{equation}
which is a Fermi distribution with chemical potential $\mu_{ij} = \mathbf{v}_i \cdot \mathbf    {v}_j$. Note that these chemical potentials are not independent, so when averaging, correlations appear when computing quantities involving three or more individuals. 

To summarize, in this letter we have examined the projection of a random bipartite network of people connected to events. Averaging over all the realizations, it is possible to extract information about the bipartite structure only with measures involving three or more agents. Indeed, lower-order quantities (e.g., the degree distribution) are not distinguishable from those of a random network if the number of events is large enough. This is evident by noting that the bipartite structure mapped in a hypercube imposes geometrical constraints. We have shown analytically how these constraints affect the properties of the projected network. In particular, we have investigated the scaling properties of the clustering coefficient in the monopartite network showing, for example, that when there are many events the system is fragmented into smaller communities.

In conclusion, our findings offer a new perspective to understanding the deep differences between monopartite and bipartite networks. 
Moreover, in most cases, we only see the monopartite network of connected individuals, and the reasons behind the observed social structure are hidden. These are too complex to trace, but our findings can provide insights into people's diversification and relationships.
Our work can offer a way to infer the underlying mechanisms (such as the co-participation in different events) giving rise to the apparent person-to-person relationship network.



\bibliography{main.bib}

\newpage
\newpage

\end{document}